# On the Lamb Vector and the Hydrodynamic Charge

**Germain Rousseaux**
Université de Nice-Sophia Antipolis,
Institut Non-Linéaire de Nice,
UMR 6618 CNRS-UNICE,
1361 route des Lucioles,
06560 Valbonne, France.

**Shahar Seifer & Victor Steinberg**
Weizmann Institute of Science,
Dept. of Physics of Complex Systems,
76100 Rehovot, Israël.

**Alexander Wiebel**
Universität Leipzig,
Institut für Informatik,
PF 100920,
D-04009 Leipzig, Germany.

**Abstract.** This work is an attempt to test the concept of hydrodynamic charge (analogous to the electric charge in Electromagnetism) in the simple case of a coherent structure such as the Burgers vortex. We provide experimental measurements of both the so-called Lamb vector and its divergence (the charge) by two-dimensional Particle Image Velocimetry. In addition, we perform a Helmholtz-Hodge decomposition of the Lamb vector in order to explore its topological features. We compare the charge with the well-known Q criterion in order to assess its interest in detecting and characterizing coherent structure. Usefulness of this concept in studies of vortex dynamics is demonstrated.

# 1. Introduction

One of the crucial issues in Turbulence is related to the identification of the so-called coherent structures in complex flows. These structures are characterized by a spatial concentration in vorticity, a life time much longer than the period of rotation (or turn-over time) and by their unpredictability. Several criteria to detect coherent structures (different from the vorticity one) were introduced in the literature and no consensus was reached so far [1]. Recently the notion of "Q criterion" (or Weiss determinant) was introduced in order to characterize turbulent flows mainly in numerical simulations: one calculates the Laplacian of the dynamic pressure alone to visualize the coherent structures [2, 3]. On the other hand, a vortex is characterized by a low pressure in its centre and its core exhibits a local velocity minimum. Indeed, the experiments of Douady & al. [4] on the visualization of coherent structures in a turbulent flow by injecting air bubbles in water show that the bubbles do accumulate in the centre of the structures that is in the pressure minima, which correspond to the vorticity maxima. These authors have also pointed out the analogy between the Q criterion and the electric charge.

The goal of our paper is to validate experimentally a criterion based on the so-called "hydrodynamic charge" with the help of an analogy between Fluid Mechanics and Electromagnetism dating back to Maxwell, to compare it with the Q-criterion, and to demonstrate the usefulness of the new criterion to study vortex dynamics. In particular, we aim at demonstrating the usefulness of this concept associated to the mapping of the so-called "Lamb vector" in order to characterize the vortex core and structure. According to our knowledge, there are almost no experimental measurements of the Lamb vector as well as the hydrodynamic charge.

## 2. The Lamb vector and the hydrodynamic charge

a. <u>Theoretical Background</u>

The Navier-Stokes equation for incompressible flow ($\nabla.\mathbf{u} = 0$) can be written under a peculiar form known as the Lamb formulation [5] :

$$\partial_t \mathbf{u} = -\nabla(\frac{p}{\rho} + \frac{\mathbf{u}^2}{2}) - 2\Omega \times \mathbf{u} - \nu\nabla \times \mathbf{w},$$

where $\Omega = \mathbf{w}/2$ is the vortex vector ($\mathbf{w} = \nabla \times \mathbf{u}$ is the vorticity) which is analogous to the angular velocity of solid rotation in solid mechanics. The Lamb vector $\mathbf{l} = \mathbf{w} \times \mathbf{u} = 2\Omega \times \mathbf{u}$ (we can find also the expression "vortex force" in the literature to denote this vector) represents the Coriolis acceleration of a velocity field under the effect of its own rotation (usually for solid body rotation, the Coriolis force is built with a rotation vector which is independent of the velocity).

Under the incompressibility constraint ($\nabla.\mathbf{u} = 0$), H. Marmanis and S. Shridar have proposed a set of "hydrodynamic Maxwell equations" which takes the form [6, 7] :

$$\nabla.\mathbf{w} = 0 \text{ or } \mathbf{w} = \nabla \times \mathbf{u}$$

$$\partial_t \mathbf{w} = -\nabla \times \mathbf{l} - \nu\nabla \times \nabla \times \mathbf{w} \text{ or } \mathbf{l} = -\partial_t \mathbf{u} - \nabla(\frac{p}{\rho} + \frac{\mathbf{u}^2}{2}) - \nu\nabla \times \mathbf{w}$$

$$\nabla.\mathbf{l} \equiv q_H = -\nabla^2(\frac{p}{\rho} + \frac{\mathbf{u}^2}{2})$$

where $q_H$ is the so-called "hydrodynamic charge density".

The first equation stands for the conservation of the vorticity flux along a tube of vorticity. It implies the absence of monopole source of vorticity as a vorticity tube that wraps on itself, goes to infinity or is connected to the walls of the flow volume. One finds the second and third equations by taking respectively the curl and divergence operators of the Navier-Stokes equation. The second one illustrates that the temporal variation of the angular velocity is equal to the torque exerted by the Coriolis force that is the Lamb vector. The third one shows that the sources of the Lamb vector are the pressure and velocity gradients.

The hydrodynamic charge can be seen as a topological feature of flow. Indeed, it is expressed mathematically as the curvature of the sum of potential (pressure) and kinetic energies. One notices that it vanishes when the flow is irrotationnal ($\mathbf{w}=\mathbf{0}$ implies $\mathbf{l}=\mathbf{0}$).

There exist localized topological structures associated with a charge that correspond to vorticity filaments, vortices, and, more generally, to the so-called coherent structures.

One can resume the analogy due to H. Marmanis and S. Shridar [6, 7] with a table. Each electromagnetic property has an hydrodynamic counterpart :

| Hydrodynamical Quantities | Electromagnetic Quantities |
|---|---|
| Specific Enthalpy $\dfrac{p}{\rho}$ | Scalar Potential $V$ |
| Velocity $\mathbf{u}$ | Vector Potential $\mathbf{A}$ |
| Vorticity $\mathbf{w}$ | Magnetic Induction $\mathbf{B}$ |
| Lamb Vector $\mathbf{l}$ | Electric Field $\mathbf{E}$ |
| Hydrodynamic Charge $q_H$ | Electric Charge $q_E$ |

The Lamb vector can be seen as a « motional » electric field analogous to the so-called Lorentz electric field which appears during a galilean transformation ($\mathbf{E'} = \mathbf{E} + \mathbf{v} \times \mathbf{B}$ and $\mathbf{B'} = \mathbf{B}$ are analogous to $\mathbf{l'} = \mathbf{l} + \mathbf{w} \times \mathbf{v}$ and $\mathbf{w'} = \mathbf{w}$). Moreover, the Lamb vector of a cylindrical vortex is radial like the electric field associated to a charged wire. That's why a vortex is a dual electromagnetic object as it is a tube of "magnetic induction", which carries an "electric charge".

One can separate the Lamb vector in two parts according to the so-called Helmholtz-Hodge decomposition [8, 9] :
$$\mathbf{l} = \mathbf{l}_{//} + \mathbf{l}_{\perp} = \nabla \alpha + \nabla \times \boldsymbol{\beta} \text{ with } \nabla . \boldsymbol{\beta} = 0$$
The indexes $//$ and $\perp$ correspond to the projections of the Lamb vector parallel (irrotationnal part) and perpendicular (solenoidal part) to the wave vector in the associated Fourier space. One infers that the hydrodynamic charge is a function only of the parallel part $\mathbf{l}_{//}$ and the scalar $\alpha$ is completely determined by the incompressibility constraint ($\nabla . \mathbf{u} = 0$) :
$$\nabla . \mathbf{l} = \nabla . \mathbf{l}_{//} = -\nabla^2 (\frac{p}{\rho} + \frac{\mathbf{u}^2}{2}) = \nabla^2 \alpha = q_H \text{ where } \alpha = -(\frac{p}{\rho} + \frac{\mathbf{u}^2}{2})$$
We notice now that the Navier-Stokes equation can be split in two parts:

$$\mathbf{l}_{//} = -\nabla(\frac{p}{\rho} + \frac{\mathbf{u}^2}{2}) \text{ and } \mathbf{l}_\perp = -\partial_t \mathbf{u} - \nu \nabla \times \nabla \times \mathbf{u}$$

The Helmholtz-Hodge decomposition of the Lamb vector is not, in general, unique as one can subtract simultaneously from $\mathbf{l}_{//}$ ($\mathbf{E}_{//}$) and add to $\mathbf{l}_\perp$ ($\mathbf{E}_\perp$) the gradient of a function, which is solution of a Laplace equation (see the numerical implementation below). For example, the vector potential (velocity) outside a solenoid (a Rankine vortex) derives from a harmonic function. Hence, its time derivative is a gradient that can enter into the longitudinal part. The boundary conditions are essential in order to guaranty uniqueness.

In a particular case of a stationary flow and when the viscous effects are negligible, the Navier-Stokes equations resume to the "auto-strophic" equilibrium (one speaks of geo-strophic equilibrium when the Earth Coriolis effect equilibrates the pressure gradient) :

$$\mathbf{l} = \mathbf{w} \times \mathbf{u} = \mathbf{l}_{//} = \nabla \alpha$$

The Lamb vector is perpendicular to the surface of constant $\alpha$ and each of these surfaces features the streamlines and the vorticity lines [10]: $\mathbf{w}.\nabla \alpha \equiv 0$ and $\mathbf{u}.\nabla \alpha \equiv 0$.

Hence, the transverse component of the Lamb vector is equal to zero and this implies the existence of a relation between the stream function and the vorticity, which in 2D is equal to minus the Laplacian of the stream function [5]. This property is very important in 2D Turbulence, which features an inverse energy cascade from the small to the large scales of the flow. Indeed, in 3D turbulence, it is the transverse component of the Lamb vector, which is at the origin of the direct cascade of energy from the large to the small scales of the flow [11, 12]. It can be shown that the longitudinal component is passive: its evolution is controlled by the transverse component [12]. In general, the transverse part is highly reduced in the region of space where the vorticity is concentrated [13].

b. <u>The Burgers Vortex</u>

The Burgers (Gaussian) vortex is a convenient model to derive the hydrodynamic charge of a typical single vortex. For comparison, in a Rankine vortex the Lamb field is discontinuous at the edge of the core and the charge diverges. We assume open cell, meaning that the characteristic radius of the Burgers vortex core, r=a, is much smaller than the cell radius. If $\Omega$ denotes the angular speed parameter of the vortex, the fields are expressed as follows ($\nu$ is the kinematic viscosity) [9]:

$$\mathbf{u}.\mathbf{e}_r = -\frac{2\nu}{a^2} r$$

$$\mathbf{u}.\mathbf{e}_\theta = \Omega \frac{a^2}{r} \left(1 - e^{-\frac{r^2}{a^2}}\right)$$

$$\mathbf{u}.\mathbf{e}_z = \frac{4\nu}{a^2} z$$

$$\mathbf{w} = 2\Omega e^{-r^2/a^2} \mathbf{e}_z$$

$$\mathbf{l} = \mathbf{l}_{//} + \mathbf{l}_\perp = w\mathbf{e}_z \times (u_r \mathbf{e}_r + u_\theta \mathbf{e}_\theta) = -w u_\theta \mathbf{e}_r + w u_r \mathbf{e}_\theta$$

$$q_H = \nabla.\mathbf{l} = -4\Omega^2 e^{-r^2/a^2} (2 e^{-r^2/a^2} - 1)$$

The hydrodynamic charge is negative except around the edge of the core at r=a. For comparison, the Rankine vortex has a rigid rotation core inside which $\nabla \cdot \mathbf{l} = -4\Omega^2$ and at the edge of the core there is an opposite delta function charge distribution. In any case, one expects the total charge per unit height ($\Delta z = 1$) to vanish:

$$\Delta q = \int_0^\infty q_H 2\pi r dr = 0$$

In the rest of the paper we study experimentally a Burgers vortex. From Particles Images Velocimetry measurements we extract the Lamb vector field and compute the hydrodynamic charge. Then, we implement a numerical Helmholtz-Hodge decomposition of the associated Lamb vector field.

## 3. Experimental Measurements

The setup consists of a single stirring rod of 40mm diameter, 110mm in length, rotating at the bottom of a water cell. The cell is cylindrical with 290mm inside diameter and 350mm height, and closed from above by a transparent Perspex plate in complete contact with the water. The tip of the rod is trimmed as a four-blade stirrer extending to the diameter of the rod where the blades are flat 3mm thick and 20mm high (see Fig. 1). The measured flow was produced with rod rotation speed of 7 rounds per second.

The flow is measured by Particle Image Velocimetry (PIV) at the plane 45mm above the rotating tip using 2mm thick light sheet from a double pulse laser (SoloPIV 532nm from New-Wave Research). The separation time between pulses is fixed at 1 ms. A double frame camera of 1 Mega-pixels is located above the cell. The analysis program is based on multipass correlation algorithm, written by Enrico Segre. The resulting instantaneous velocity field is determined on 80X80 vector grid. The velocity fields were acquired at a rate of 4 Hz during 38 seconds.

From the PIV measurements, one infers the velocity field and computes the radial profile of the azimuthal velocity field (Figs. 2 a-b). The fitted parameters according to the Burgers model are $\Omega = 86$ rad/s, and $a = 9.2$ mm. For all plots, the scaling is linear and the pictures are averaged over all acquired images. Without averaging, the plots are very similar. We used averaging in order to smooth the data for computation purposes especially for the Helmholtz decomposition. The error bars in Figs.2-b,d,f are based on comparison with similar results of another set of measurements (performed at 480rpm rotation speed, with velocities scaled according to the rotation speed difference by 540/480).The vertical vorticity shown in Figures 2 c-d is deduced from the velocity field. The hydrodynamic charge is displayed in Figures 2 e-f. One notices that the negative core is surrounded by a positive annulus of charge that decays farther outside to zero similarly to the Lamb vector. The radial profiles of the vorticity and the charge compare reasonably with the calculations of the Burgers vortex with the same parameters as above. The apparent deviations between experiment and theory in Figure 2f can be explained by viscous effects, which tend to smooth out the velocity profile hence all derived quantities like the Lamb vector and the charge especially at the outer edge of the vortex where the velocity reaches a maximum. Moreover, the center of the vortex is not fixed so in each frame the radial profile is determined with different reference point. Possibly, the advection of vorticity is not left without trace and part of the coherence of the structure is destroyed either by the sampling average or by a dynamical mechanism in the flow. This means that the local charge is smoothed in the center and does not display an ideal vortex structure.

## 4. Discrete Helmholtz-Hodge Decomposition

As mentioned above, we use the Helmholtz-Hodge decomposition to separate the Lamb vector into an irrotational and a solenoidal part. The decomposition is implemented in a way similar to Tong & al. [14].

The discrete Helmholtz-Hodge decomposition of the Lamb vector $l$ tries to mimic the smooth decomposition described by $l = d + r + h = \nabla D + \nabla \times R + h$. Here $d$ and $r$ are the irrotational and the solenoidal part respectively. The so-called harmonic part $h$ is both irrotational and solenoidal. As in 2D the curl is not a vector we have to rewrite the decomposition for 2D as $l = d + r + h = \nabla D + J\nabla R + h$ where $J\nabla$ is the co-gradient and $J$ rotates every vector by ninety degrees in counter clock-wise order (see Polthier & Preuß [15]). One easily verifies that the divergence of $r = J\nabla R$ vanishes and thus $r$ is solenoidal in the two dimensional case too. The harmonic part can be added to $d$ or $r$ without changing their main characteristic (solenoidal / irrotational). Thus the decomposition can be written as previously : $l = d' + r' = \nabla \alpha + \nabla \times \beta$.

In the smooth case $\nabla D$ and $\nabla \times R$ correspond to projections of $l$ onto the spaces of curl-free resp. divergence-free vector fields. Such projections can be achieved by minimizing the following functionals:

$$F(D) = \frac{1}{2}\int_T (\nabla D - l)^2 dV$$

$$G(R) = \frac{1}{2}\int_T (J\nabla R - l)^2 dV$$

Trying to mimic the smooth case, the two potentials $D$ and $R$ are derived by projections in the discrete case too; see Tong & al. [14] for details. Discretization of the above equations leads to a sparse linear system for each potential. To guarantee unique solutions boundary conditions have to be specified. We follow Tong and choose the boundaries of the potentials to be zero. With this choice $d$ is orthogonal to the boundary and $r$ is tangential to it.[1] To solve the sparse systems, we employ a standard conjugate gradient method. The solution vectors contain the values of the potentials for each point. Computing the gradient and the co-gradient respectively yields the desired components of the field $l$.

Since the decomposition is a global variational approach and as the potentials are computed by integrated values the decomposition has smoothing effects on the results. This is valuable as we are working with measured data.

For all plots in Figure 3, the streamlines associated to each type of vector are started at randomly distributed points. The Lamb streamlines are emitted from the centre of the vortex. If the Lamb vector was only a gradient the streamlines should be star-like. However, we notice for our experimental vortex a small spiralling behaviour. Hence, we infer the presence of, at least, a transverse part of the Lamb vector, which indeed looks like a four-leaved clover, which reminds us the symmetry of the stirrer (Fig. 1). In addition, both the divergence-free and the harmonic part of the Lamb vector are asymmetric.

---

[1] To avoid confusion, note that all figures show only a cut-out of the whole field. The boundary is not shown.

The spiralling behaviour is due to the presence of the radial and vertical components of the velocity, which do create a transverse component of the Lamb vector. Quoting Saffman ([16], p. 47): " If in a steady flow, $\mathbf{l}$ is not the gradient of a single valued scalar, which can be absorbed in the pressure, then an external body force must be applied to maintain equilibrium". In the present experiment this is the spinning axis and propeller. The motion is accompanied by a radial and an axial velocity. We checked experimentally using Burgers model that the azimuthal component of the Lamb vector is very small compared to its radial part as the former one is proportionnal to the kinematic viscosity:

$$\mathbf{l}_{//} = -\frac{2\Omega^2 a^2}{r} e^{-r^2/a^2} \left(1 - e^{-r^2/a^2}\right) \mathbf{e}_r \gg \mathbf{l}_\perp = -\frac{4\nu\Omega}{a^2} e^{-r^2/a^2} r \mathbf{e}_\theta$$

## 5. Criteria for detection of the vortex centre.

In order to compare the criterion based on the Lamb vector and hydrodynamic charge (see Fig.2e-f) with one of the considered before criteria, namely Q-criterion, let us first relate these two criteria. Indeed, the hydrodynamic charge can be easily computed with the gradient of the kinetic energy and the Q-criterion via expressions:

$$\nabla \cdot \mathbf{l} = -\nabla^2 \left(\frac{p}{\rho} + \frac{\mathbf{u}^2}{2}\right) = -2Q - \nabla^2 \left(\frac{\mathbf{u}^2}{2}\right) = q_H$$

hence :

$$Q = -\frac{1}{2}\left(q_H + \nabla^2 \left(\frac{\mathbf{u}^2}{2}\right)\right)$$

The vortex cores are usually characterized by strong positive values of Q (as the pressure reaches a minimum in the centre) and we can identify the vortex cores as the circular regions with positive Q around a peak of vorticity. As discussed previously, the hydrodynamic charge takes into account both the influence of pressure (Q) and the influence of velocity. Indeed, the vortex centre corresponds to a local minimum of these two quantities. The Q-criterion captures only minima of pressure. Hence, the hydrodynamic charge is a kind of modified Q-criterion that takes into account also influence of the kinetic energy.

In Fig. 4a, we superimposed visualizations of the Lamb vector and the hydrodynamic charge fields. This image actually reproduces the same data and in the same form that shown in Fig.2e but obtained by different numerical procedure via discrete Helmholtz-Hodge decomposition. Figure 4b presents the following data: the color and color-bar visualize field of div(grad($\mathbf{u}$/2)), the scaled arrows show grad($\mathbf{u}$/2), the line integral convolution (LIC) in the background shows the structure of grad($\mathbf{u}$/2), the dots with the white lines show the topology of grad($\mathbf{u}$/2). The red squares are saddle points and the blue squares are sinks. Both are zeros of the vector field displayed. In Fig. 4c the Q-criterion is presented. One can learn from the images that the hydrodynamic charge and the Lamb vector fields characterize the vortex in the clearest way.

## 6. Vortex dynamics characterization

An immediate and useful application of the Lamb vector and hydrodynamic charge concepts is to study dynamics of coherent structures in turbulent flow. As an example we examined first movies of PIV images by using the divergence-free (solenoidal) part and rotational-free part of the Lamb vector field and found a clear episode pattern. Ordinarily in the rotational-free part we observe a cloud of vectors pointing towards the centre of the vortex (the direction does not change in either rotation direction of the vortex). Only occasionally we observe events that the vortex breaks into separate hydrodynamic charges. A cloud emerges with outward pointing vectors beside the inward pointing vectors cloud, but in the next frame the ordinary situation is recovered (see Fig. 5). In the solenoidal picture of such events we observe disturbances with high magnitudes.

In order to identify numerically the occurrence of special events we describe two results of analysis by intuitive names as the "polarity" and the "EMF" (electro-motive force). The polarity is a quantity of the maximal positive charge minus the maximal negative charge observed in the frame (the charge is the divergence of the rotational-free part of the Lamb vector). The EMF is the net value of the solenoidal part, presented as the magnitude of the averaged vector over the entire frame. As we have shown for a single eddy, the Lamb vector in solenoidal part is related to the time derivative of the global vorticity (that is analogous to a time derivative of a magnetic flux that produces EMF). Accordingly in Fig. 6 we show the trajectory of the vortex centre in x-y coordinate, which determination is based on the Lamb vector and hydrodynamic charge criterion, the polarity and the EMF (in arbitrary units) versus time. Clearly, there are specific events that are identified by peaks in both the polarity and EMF plots. For example, the episode shown in Fig. 5 relates to the first peak at t=0.75 sec. In addition, there is a correlation between the peaks and abrupt changes in the course of the vortex trajectory, as marked by arrows in Fig. 6. One can also notice that large EMF peaks predict large instability of the trajectory (indicated by larger displacements). This analysis demonstrates usefulness of the approach to identify the vortex core and to provide detail information about dynamics of coherent structures that has been unavailable before.

## 7. Conclusions and Perspectives

In this paper we demonstrate the usefulness and effectiveness of the concepts of the Lamb vector and the hydrodynamic charge to locate and characterize coherent structures such as a vortex on the experimental data. In addition, the topology of the Lamb vector seems also to be a rather robust indicator for the presence of coherent structures particularly by applying the Helmholtz-Hodge decomposition.

The analogy with electromagnetism is a useful guide but should not be extended too much. For example, as the Navier-Stokes equations are non-linear contrary to Maxwell equations, there is no such thing as stretching in Electromagnetism. Still, the notion of hydrodynamic charge has some clear soundness…

However, the scope of our results is limited as one needs to confirm such indications for more complicated situations such as described in the numerical simulations of Kollmann and Umont [17]. From the experimental point of view, three-dimensional PIV seems to be a good candidate in order to explore more deeply the role of the Lamb vector and its divergence in our understanding of coherent structures in turbulence.

## 8. Acknowledgement

Two of us (Sh. S. and V. S.) are grateful to E. Segre for providing us a multi-pass correlation algorithm and for his help in software support. This work is partially supported by grants from Israel Science Foundation, Binational US-Israel Foundation, and by the Minerva Centre for Nonlinear Physics of Complex Systems. G.R. was financially supported by a grant "post-doc CNRS" (S.P.M. section 02) during his post-doctoral stay in Nice. A.W. was supported by DFG grant SCHE 663/3-7.

Figure 1: (Experimental Vortex) A snapshot of the divergence-free part of the Lamb vector and the rotating rod tip that may explain the peculiar symmetry.

Figure 2 (Colour Online) : (Experimental Vortex) a. velocity field and its LIC visualisation (LIC is the line integral convolution), b. radial profile of the azimuthal velocity, c. vertical vorticity (in $s^{-1}$), d. radial profile of the vertical vorticity, e. planar Lamb vector and planar hydrodynamic charge (in $s^{-2}$), f. radial profile of the hydrodynamic charge. The profile fits are based on the Burger model. Error bars are displayed in red.

Figure 3 : (Experimental Vortex) a. planar Lamb vector field and its streamlines, b. divergence-free part of the Lamb vector, c. vorticity-free part of the Lamb vector and d. harmonic part of the Lamb vector.

Figure 4 (Color Online): (Experimental Vortex) a. hydrodynamic charge, b. gradient of the kinetic energy, c. Q criterion = Laplacian of the pressure.

Figure 5: (Experimental Vortex) A typical episode of spontaneous disturbances in the vortex shape and course as observed in the irrotational part of the Lamb vector (time t=0.5s; 0.75s; 1s).

Figure 6 (Color Online): (Experimental Vortex) Summary of correlated quantities in vortex dynamics for spontaneous disturbances: the centre position of the vortex, the polarity, and EMF versus time (see definitions in text). Arrows indicate the point of the event on the 3D plot of the trajectory.

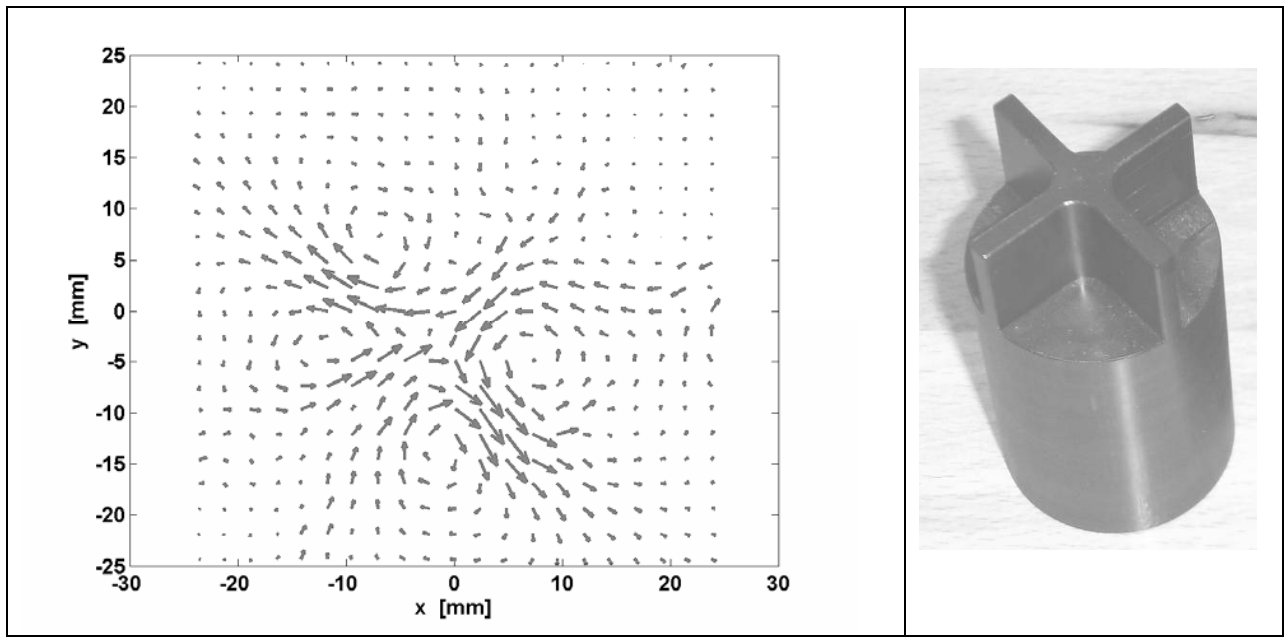

Figure 1

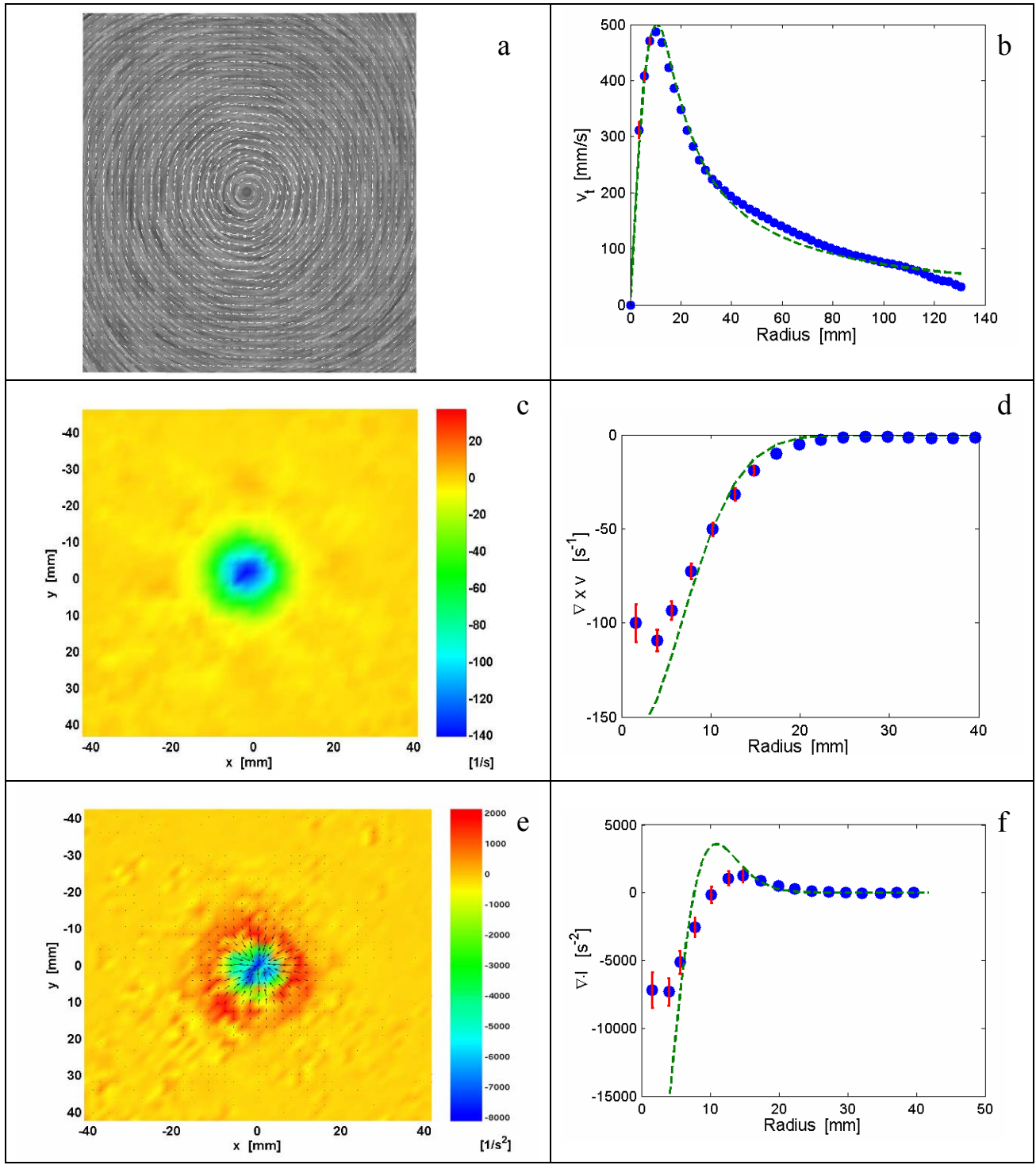

Figure 2 (Color Online)

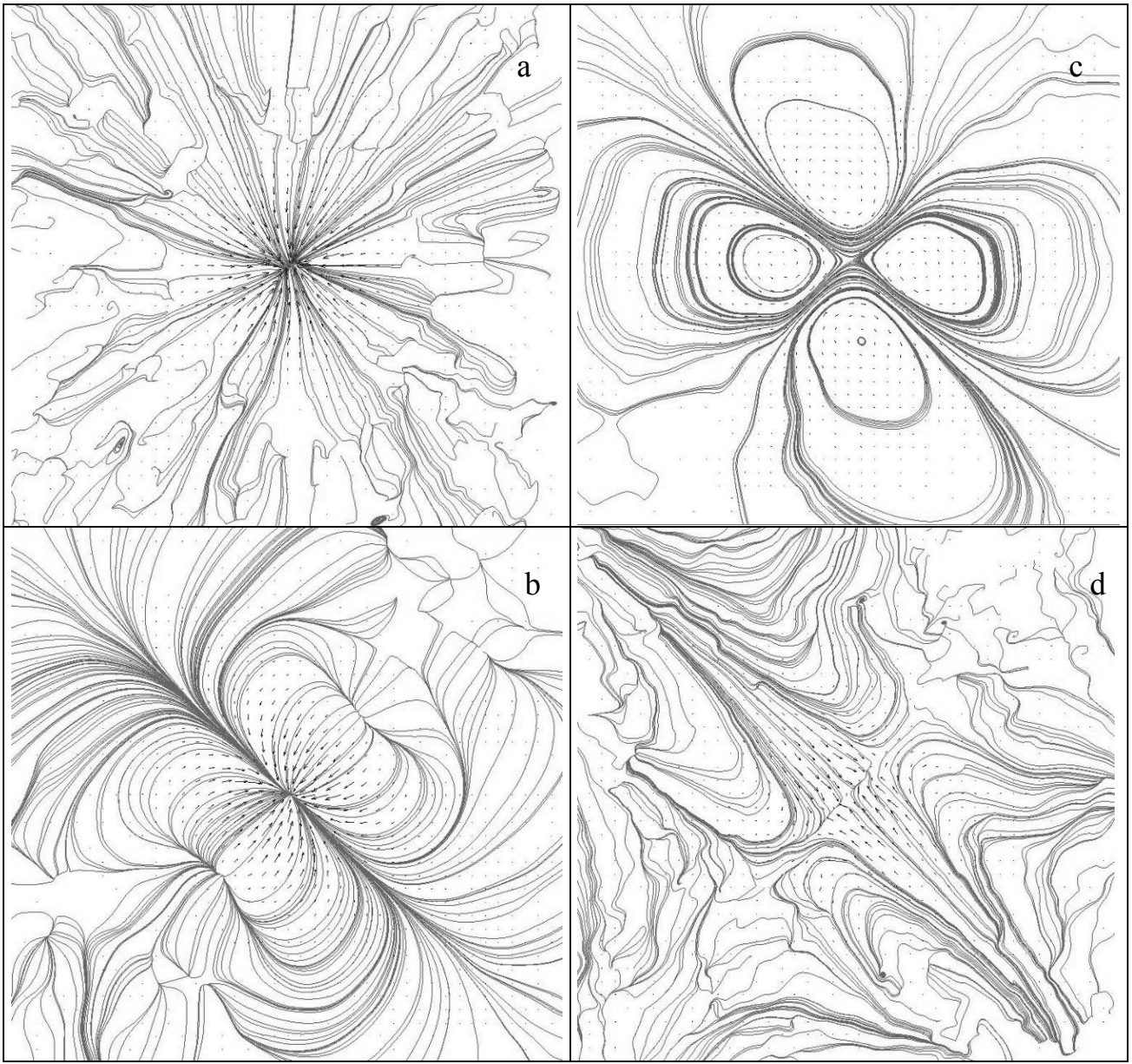

Figure 3

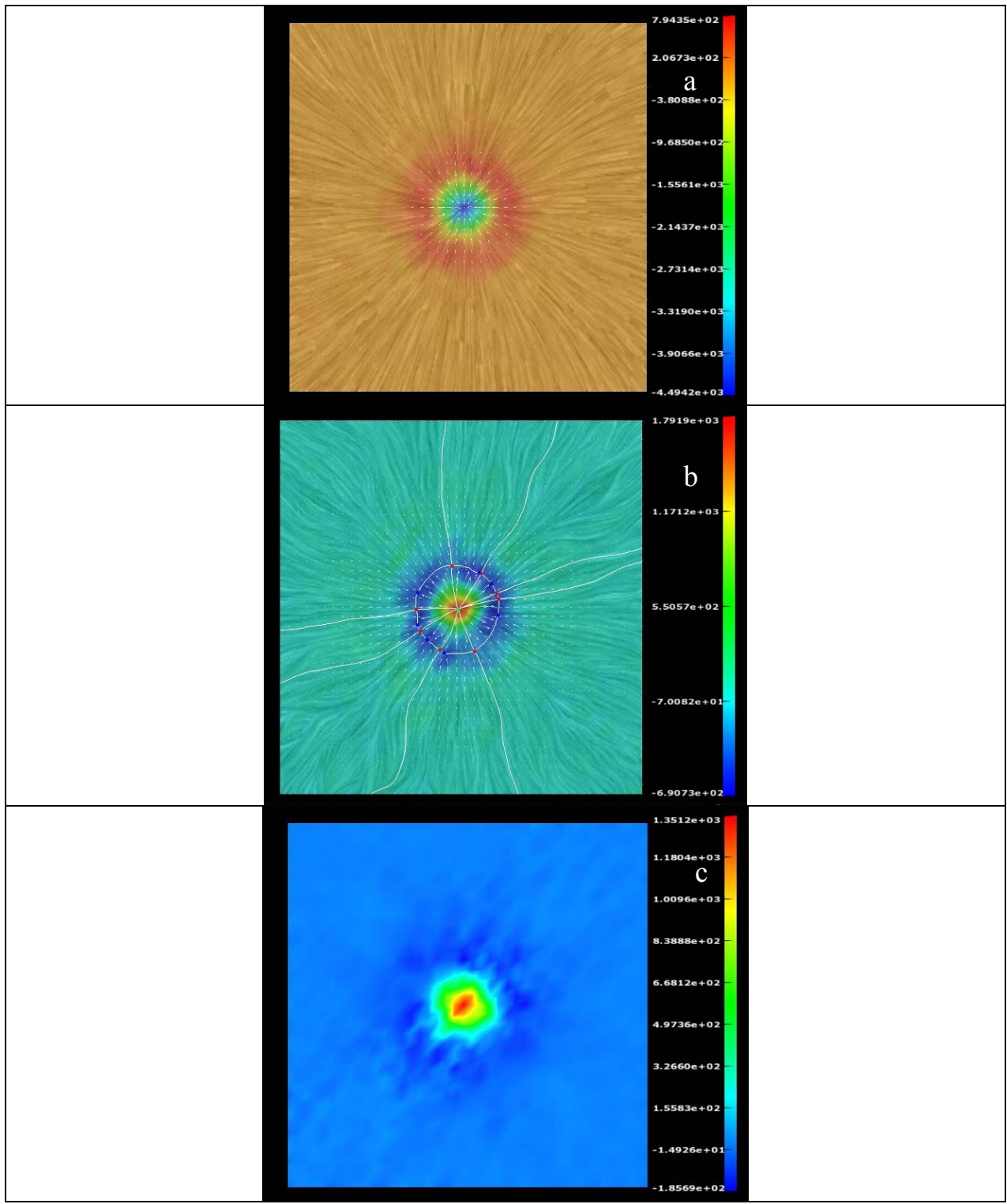

Figure 4 (Color Online)

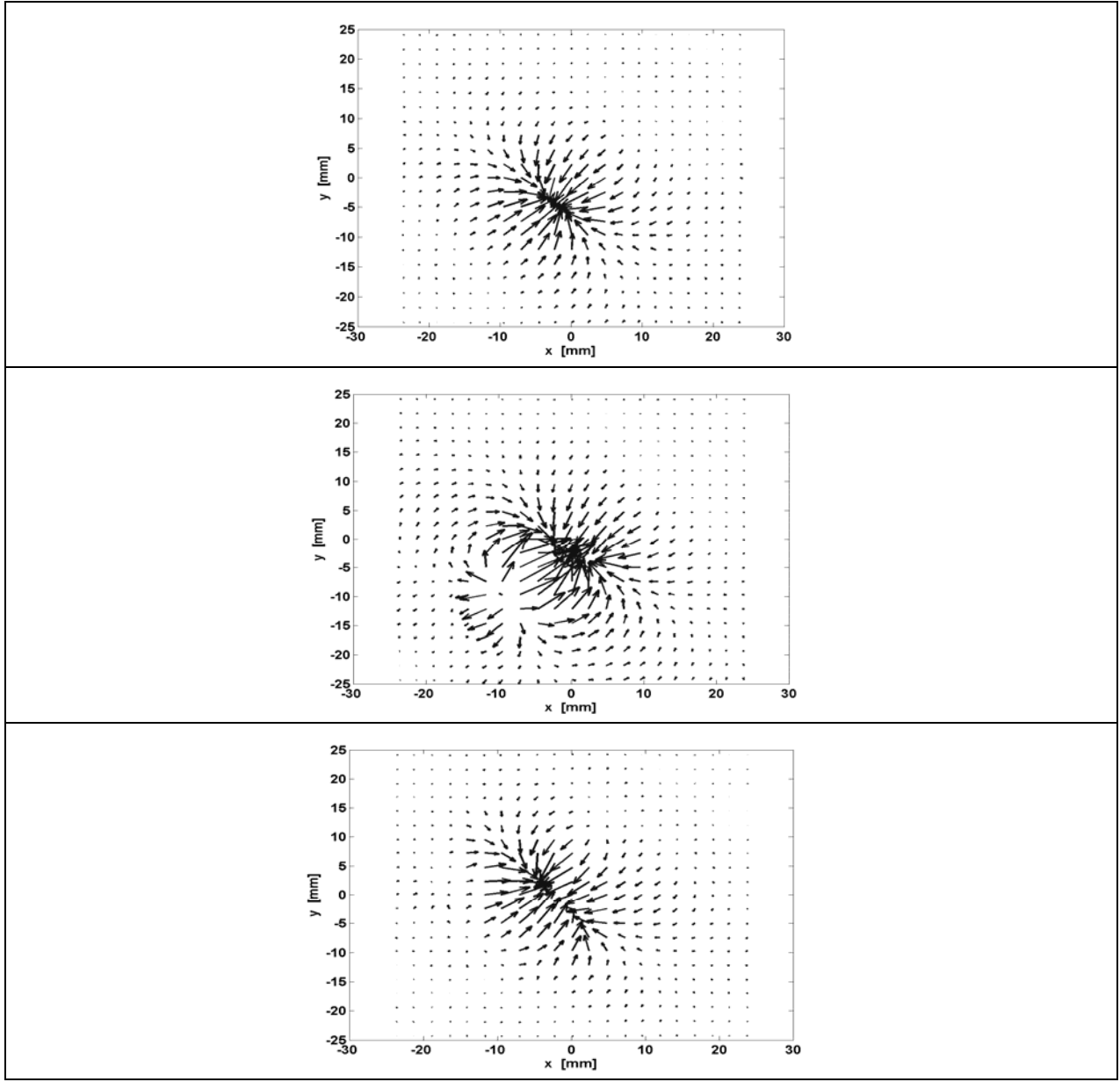

Figure 5

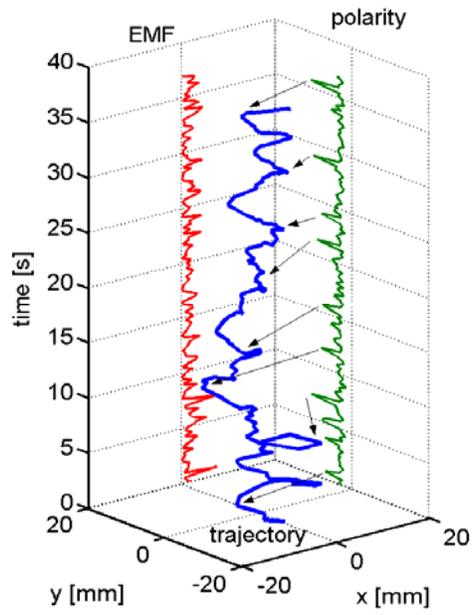

Figure 6 (Color Online)